\documentclass[9pt,twocolumn,twoside]{pnas-new}

\templatetype{pnasresearcharticle} 

\title{Computation of extreme heat waves in climate models using a large
deviation algorithm}

\author[a,b]{Francesco Ragone}
\author[a,c,d]{Jeroen Wouters} 
\author[a]{Freddy Bouchet}

\affil[a]{Laboratoire de Physique, Ens de Lyon, Univ Claude
Bernard, Univ Lyon, CNRS, F-69342 Lyon, France}
\affil[b]{Department of Earth and Environmental Sciences,
University of Milano\textendash Bicocca, Milan, Italy}
\affil[c]{School of Mathematics and Statistics, University
of Sydney, Sydney, Australia}
\affil[d]{Meteorological Institute, University of Hamburg,
Hamburg, Germany}

\leadauthor{Ragone} 

\significancestatement{We propose an algorithm to sample rare events in climate models that costs hundred to thousand times less than direct sampling of the model. Applied to the study of extreme heat waves, we estimate the probability of events that can not be studied otherwise because they are too rare, and we get a huge ensemble of realizations of extreme event. Using these results, we describe the teleconnection pattern for the extreme European heat waves. This method should change the paradigm for the study of extreme events in climate models: it will allow to study extremes with higher complexity models, to make inter model comparison easier, and to study the dynamics of extreme events with an unprecedented statistics.}

\authorcontributions{F.B. proposed the project, initiated and directed the work; F.R., J.W., and F.B. performed
	research; F.R. made the core of the numerical work that led to the rare event computation of heat waves and the resulting analysis; and F.R., J.W., and F.B. wrote the paper.}
\authordeclaration{}
\correspondingauthor{\textsuperscript{1}To whom correspondence should be addressed. Freddy.Bouchet@ens-lyon.fr}


\begin{abstract}
Studying extreme events and how they evolve in a changing climate
is one of the most important current scientific challenges. Starting
from complex climate models, a key difficulty is to be able to run
long enough simulations in order to observe those extremely rare events.
In physics, chemistry, and biology, rare event algorithms have recently
been developed to compute probabilities of events that cannot be observed
in direct numerical simulations. Here we propose such an algorithm,
specifically designed for extreme heat or cold waves, based on statistical
physics. This approach gives an improvement of more than two orders
of magnitude in the sampling efficiency. We describe the dynamics
of events that would not be observed otherwise. We show that European
extreme heat waves are related to a global teleconnection pattern
involving North America and Asia. This tool opens up a wide range of
new possible studies to quantitatively assess the impact of climate
change.
\end{abstract}

\dates{This manuscript was compiled on \today}
\doi{\url{www.pnas.org/cgi/doi/10.1073/pnas.1712645115}}

\begin{document}


\maketitle
\thispagestyle{firststyle}
\ifthenelse{\boolean{shortarticle}}{\ifthenelse{\boolean{singlecolumn}}{\abscontentformatted}{\abscontent}}{}

\dropcap{R}are events, for instance extreme droughts, heat waves, rainfalls
and storms, can have a severe impact on ecosystems and socioeconomic
systems \cite{intergovernmentalpanelonclimatechange_managing_2012,aghakouchak_extremes_2012,herring_explaining}.
The Intergovernmental Panel on Climate Change (IPCC) has concluded
that strong evidence exists indicating that hot days and heavy precipitation
events have become more frequent since 1950 \cite{intergovernmentalpanelonclimatechange_climate_2013,coumou_decade_2012}.
However, the magnitude of possible future changes are still uncertain
for classes of events involving more dynamical aspects, like for instance
hurricanes or heat waves \cite{intergovernmentalpanelonclimatechange_climate_2013,Shepherd2015,shepherd2016common}.
Estimates of the average time in between two events of the same class,
called return time (or return period), is key for assessing the expected changes in extreme
events and their impact. This is crucial on a national level when
considering adaptation measures and on the international level when
designing policy to implement the Paris Agreement, in particular its
Art. 8 (\href{https://en.wikisource.org/wiki/Paris_Agreement\#Article_8}{https://en.wikisource.org/wiki/Paris\_{}Agreement}). Public or private risk managers need to know amplitudes of events with a return time ranging from a few years, to hundreds of thousands of years when the impact might be extremely large.

The 2003 Western European heat wave led to a death toll of more than
$70\,000$ \cite{robine_death_2008}. Similarly, the estimated impact
of the 2010 Russian heat wave was a death toll of 55,000, an annual
crop failure of $\sim25\%$, more than 1 million ha of burned areas,
and $\sim$US\$15 billion ($\sim1\%$ of gross domestic product) of
total economic loss \cite{barriopedro_hot_2011}. During that period,
the temperature averages over 31 days at some locations were up to
5.5 standard deviations away from the 1970-1999 climate \cite{barriopedro_hot_2011}.
As no event similar to those heat waves has been observed during the
last few centuries, no past observations exist that would allow to
quantify their return times. If return times cannot be estimated
from observations, we must rely on models. 

Several scientific barriers need to be overcomed, however, before we can obtain quantitative estimates of rare event return times from a model. One of them
is that extreme events are observed so rarely that collecting sufficient
data to study quantitatively their dynamics is prohibitively
costly. This has led authors of past studies to either use models which are of a
lesser quality than the up to date IPCC top class models, or to focus
on a single or a few events which does not allow for a quantitative
statistical assessment. Making progress for this sampling issue would
also allow a better understanding of those rare event dynamics,
and strengthen future assessment of which class of models
is suited for making quantitative predictions. 

\section*{Rare event algorithms}

In physics, chemistry, and biology rare events may matter: even if
they occur on time scales much longer than the typical dynamics time scales, they
may have a huge impact. During the last decades, new numerical tools,
specifically dedicated to the computation of rare events from the
dynamics but requiring a considerably smaller computational effort,
have been developed. They have been applied for instance to changes
of configurations in magnetic systems in situations of first order
transitions \cite{e_energy_2003,kohn_magnetic_2005,rolland2016computing}, chemical reactions
\cite{chandler_interfaces_2005}, conformal changes of polymer and
biomolecules \cite{noe_constructing_2009,metzner_illustration_2006,bolhuis_transition_2002,wolde_enhancement_1997} and
rare events in turbulent flows \cite{laurie2015computation,grafke2013instanton,grafke2014arclength,bouchet2014langevin,bouchet2017fluctuations}.
Since their appearance \cite{kahn_estimation_1951}, the analysis of these rare event
algorithms also became an active mathematical field \cite{rubino_rare_2009,bucklew_introduction_2004,delmoral_feynmankac_2004,brehier_analysis_2014,cerou_adaptive_2007}.
Several strategies prevail, for instance genetic algorithms where
an ensemble of trajectories is evolved and submitted to selections,
minimum action methods, or importance sampling approaches. 

Here we apply for the first time a rare event algorithm for sampling
extreme events in a climate model. Given the complexity of the models and phenomena, this has long been thought to be impracticable for climate applications. A key success factor for this approach is to first clearly identify a restricted class of phenomena for which rare event algorithm may be practicable. Then one has to choose among the dozens of available algorithms which one may be suited for this class of phenomena. Finally one has to develop the tools that will make one specific algorithm effective for climate observables. Matching these idea coming from the rare event community and climate dynamics requires a genuine interdisciplinary approach, in order to both master the climate dynamics phenomenology and the probability concepts related to rare event algorithms.
We study extreme heat waves as a robust phenomena in current climate
models, involving the largest scales of the turbulent dynamics, and
use an algorithm dedicated to study large deviations of time averaged
quantities: the Giardina\textendash Kurchan\textendash Lecomte\textendash Tailleur
(GKLT) algorithm \cite{giardina_direct_2006,tailleur_probing_2007,lecomte_numerical_2007}. As this algorithm was dedicated to compute large deviation rate functions in the infinite time limit, we have to pick the main ideas of the algorithm, but to twist its usage to compute finite time observables. Moreover, we have to develop a further adaptation, aimed at computing return times rather than large deviation rate functions or tails of probability distribution functions. 
With this approach, based on statistical physics concepts, we compute
the probability of events that cannot be observed directly in the
model, the number of observed rare events for a given amplitude is
multiplied by several hundreds, and we can predict the return time
for events that would require thousand times more computational resources.

\begin{figure}
\centering
(a)\includegraphics[width=0.6\linewidth]{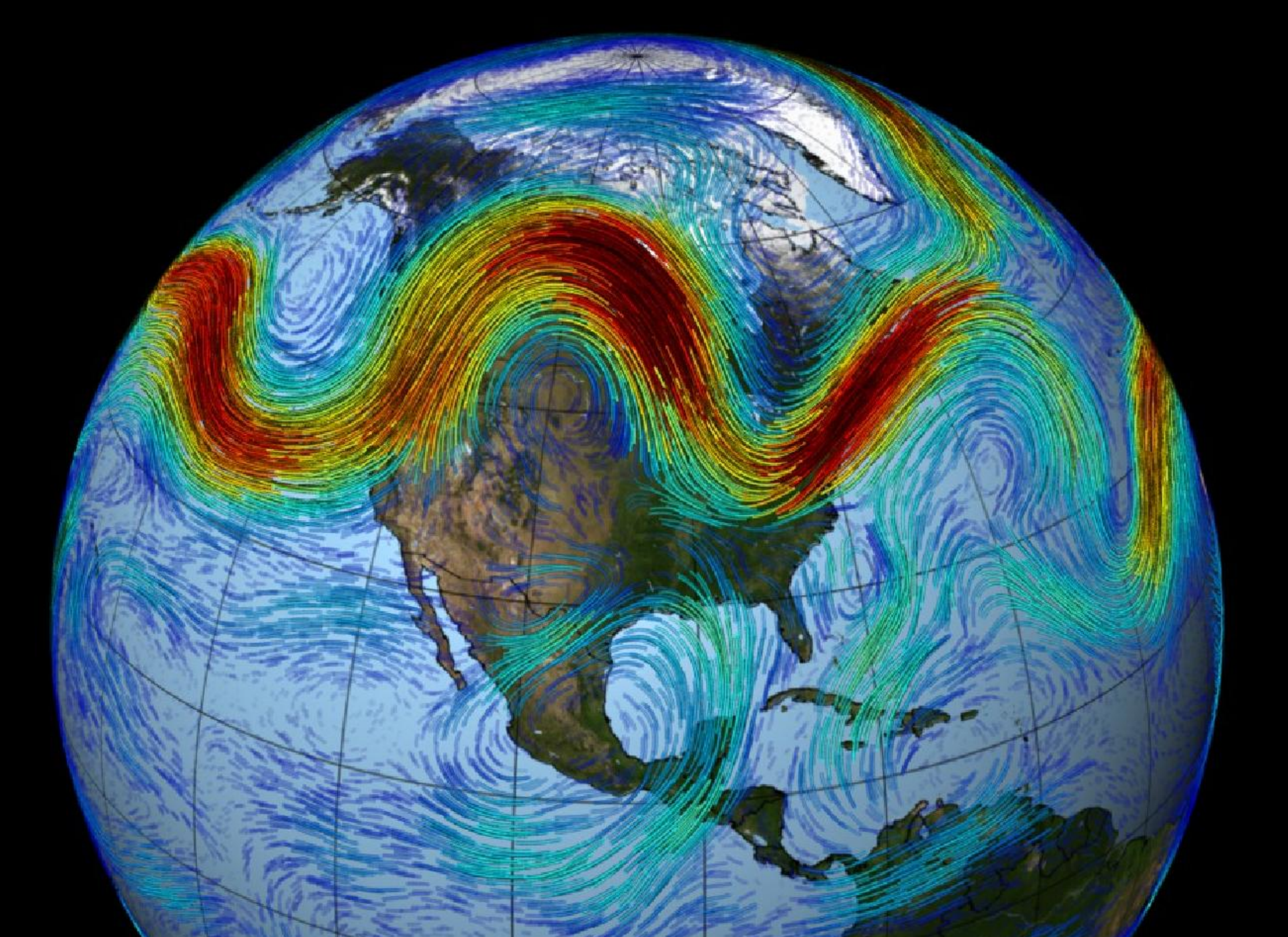}
(b)\includegraphics[width=0.6\linewidth]{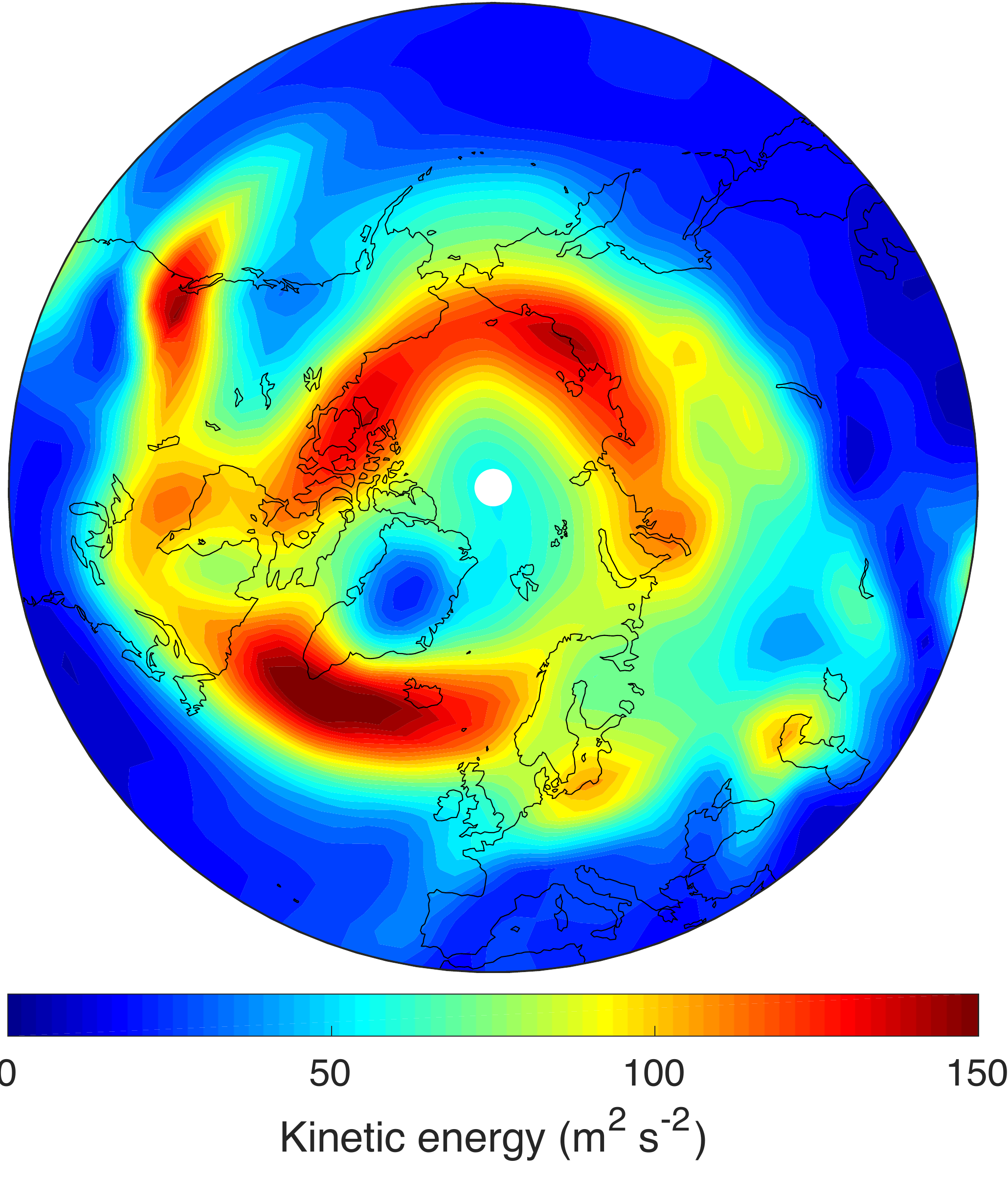}
\caption{(a) Snapshot of wind speed velocity at the top of the troposphere,
showing the jet stream over North America (from NASA). (b) Average
horizontal kinetic energy at 500hPa (mid troposphere) in the Plasim
model, showing the averaged Northern Hemisphere jet stream.}
\label{fig:Fig1}
\end{figure}

\section*{The jet stream dynamics and extreme heat waves}

Midlatitude atmospheric dynamics is dominated by the jet streams (one
per hemisphere). The jet streams are strong and narrow eastward air
currents, located at about $45\text{\textdegree\,N}$ or $45\text{\textdegree\,S}$,
with maximum velocity of the order of $40\,\text{m.\ensuremath{s^{-1}}}$
close to the tropopause (see Fig. \ref{fig:Fig1}(a)). The climatological
position of the northern hemisphere jet stream in our model is seen on Fig. \ref{fig:Fig1}(b),
that represents the time average of the kinetic energy due to the
horizontal component of the velocity field at 500 hPa pressure surfaces.
The jet stream's meandering dynamics, due to non-linear Rossby waves,
is related to the succession of anticyclonic and cyclonic anomalies
which characterize weather at midlatitudes. It is well known that
midlatitude heat waves, like the 2003 Western European heat wave or
the 2010 Russian heat waves, are due to rare and persistent anticyclonic
anomalies (or fluctuations), that arise as either Rossby wave breaking
(blockings), or shifts of the jet stream, or more complex dynamical
events leading to a stationary pattern of the jet stream. 

Studying extreme heat waves then amounts to studying the non-linear
and turbulent dynamics of the atmosphere. Two key dynamical variables
are the temperature and pressure fields. One could look at pressure
maps at some value of the geopotential height (the most convenient
vertical coordinate). Equivalently, it is customary to look at the
geopotential height value on a surface defined by a fixed pressure.

\begin{figure}
\centering
(a)\includegraphics[width=0.6\linewidth]{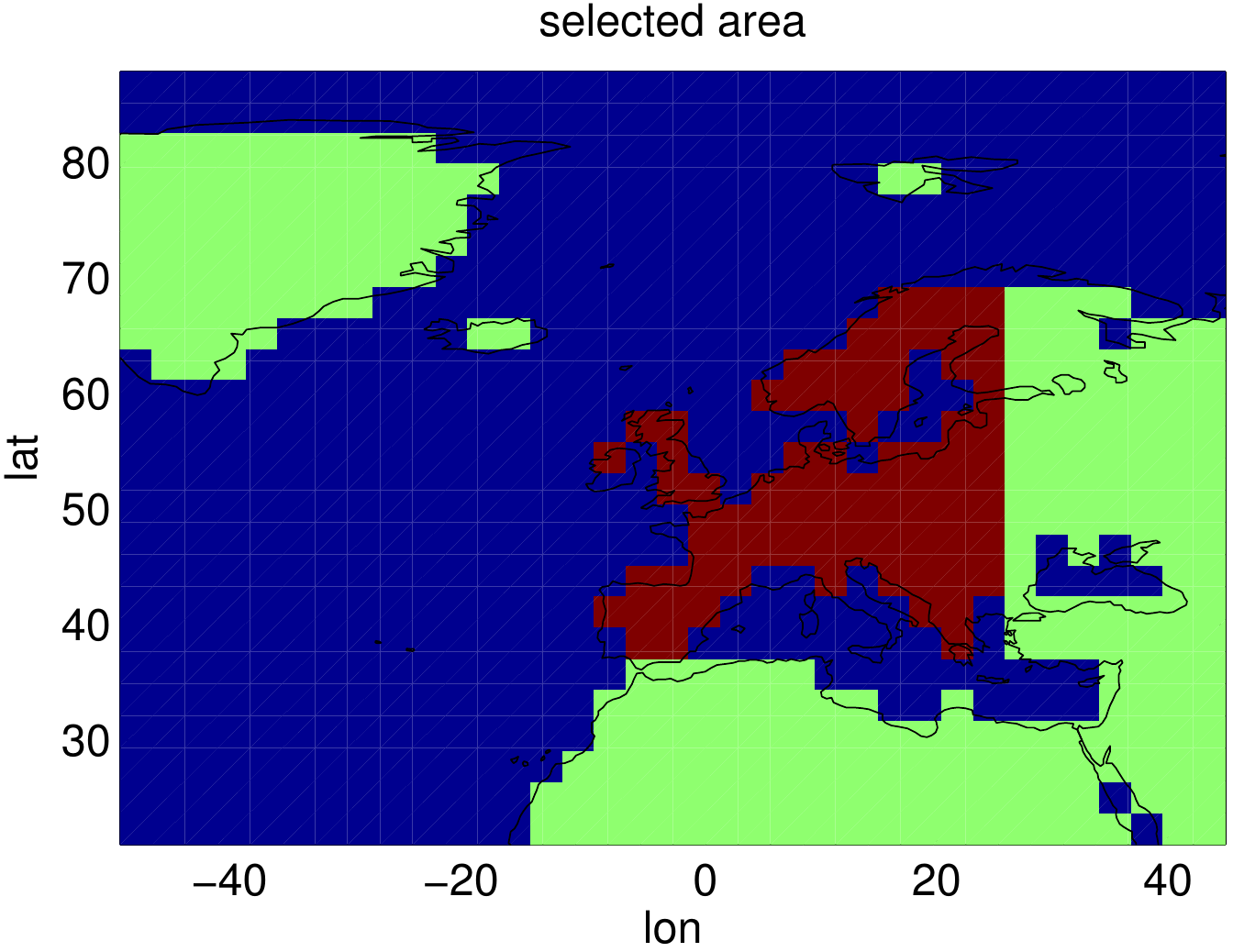}
(b)\includegraphics[width=0.6\linewidth]{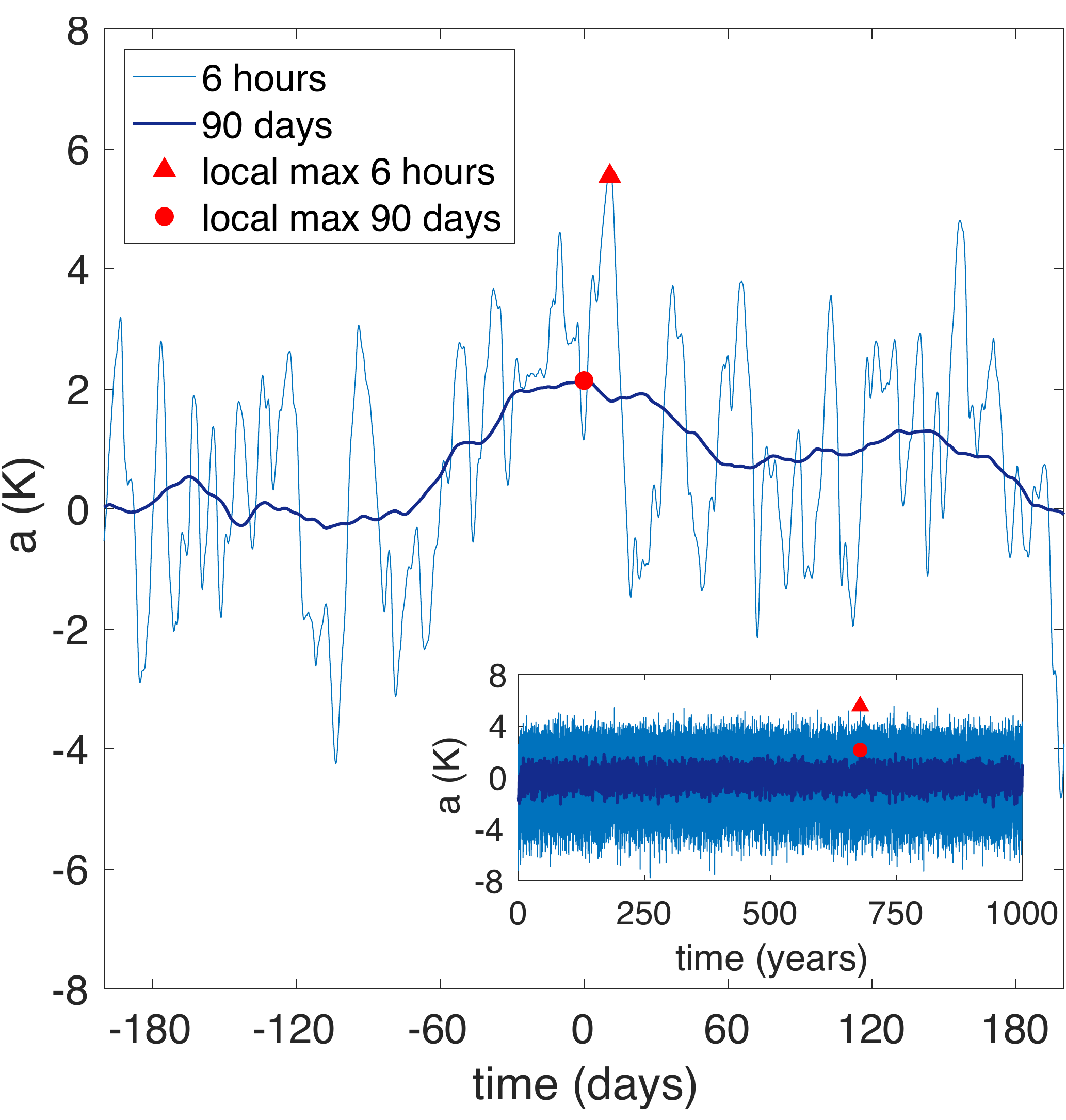}
\caption{(a) The red colors marks the Europe area over which the temperature
is averaged. (b) Timeseries of European surface temperature anomaly,
6 hours (light blue) and 90 days running mean (dark blue), during
360 days and $1\,000$ years (inset). The red triangles and circles
feature one local maximum of the temperature anomalies, as an example
of a $2\,K$ heat wave lasting $90$ days.} 
\label{fig:Fig2}
\end{figure}

\section*{Heat waves in the Plasim model}

We use the Plasim model (Planet Simulator, \cite{Fraedrich&al2005}).
Plasim gives a reasonably realistic representation of atmospheric dynamics
and of its interactions with the land surface and with the mixed layer
of the ocean; it includes parameterizations of radiative transfers
and cloud dynamics. While Plasim features about $10^{5}$ degrees
of freedoms, it is simpler and less computationally demanding than
the top class IPCC models used for assessing the projection of temperature
increase. It is nevertheless in the class of models used to discuss
extreme heat waves in the last IPCC report (for instance \cite{Otto&al2012}).
Our aim is to demonstrate the huge potential of rare event algorithms for
this class of models, and to advocate the feasibility of this approach
for top class IPCC models in the near future. 

Heat waves can be defined as rare and long lasting anomalies (fluctuations)
of the surface temperature over an extended area \cite{Stefanon&al2012,fischer_consistent_2010}.
We consider
\begin{equation}
a=\frac{1}{T}\int_{0}^{T}\tilde{A}(t)\,\mbox{d}t\,\,\,\mbox{where}\,\,\,\tilde{A}(t)=\frac{1}{\mathcal{A}}\int_{\mbox{Area}}T_{S}(\mathbf{r},t)\,\mbox{d}\mathbf{r},\label{eq:Observable-1}
\end{equation}
where $\mathbf{r}$ is the spatial variable, $t$ is time, and $T_{S}$
is the surface temperature anomaly with respect to its averaged value,
$\mathcal{A}$ is the surface area, and $T$ is the heat wave duration.\textbf{
}The relevant value for $T$ depends on the application of interest.\textbf{
}We vary $T$ between a few weeks to several months and discuss the
results for $T=90$ days. The spatial average is over Western Europe,
the region over land surface with latitudinal and longitudinal boundaries
36N-70N and 11W-25E (see Fig. \ref{fig:Fig2}(a)). We study
the upper tail of the probability distribution function (PDF) of $a$,
denoted $P(a,T)$. 

The instantaneous $T_{S}$ has standard deviation $\sigma\approx1.6$
$K$, and is slightly skewed towards positive values. Its autocorrelation
time is $\tau_{c}\approx7.5$ days, which is the typical time for
synoptic fluctuations (at a scale of about $1\,000\,km$). An example
of a timeseries of the $90$ days averaged Europe temperature anomaly
is shown on Fig. \ref{fig:Fig2}(b). 

\section*{Importance sampling and large deviations of time averaged temperature}

\begin{figure}
\centering
(a)\includegraphics[width=0.5\linewidth]{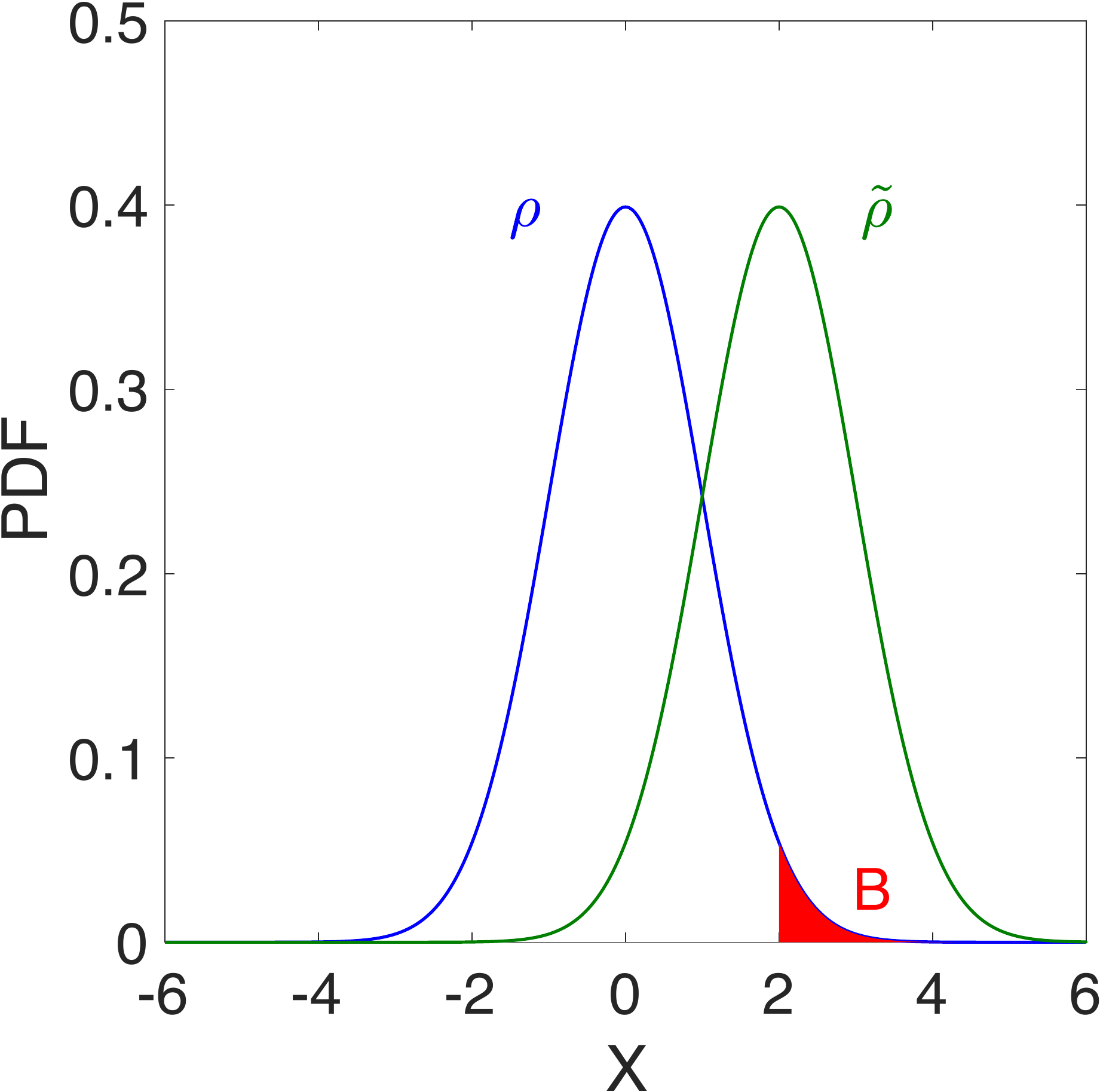}
(b)\includegraphics[width=0.5\linewidth]{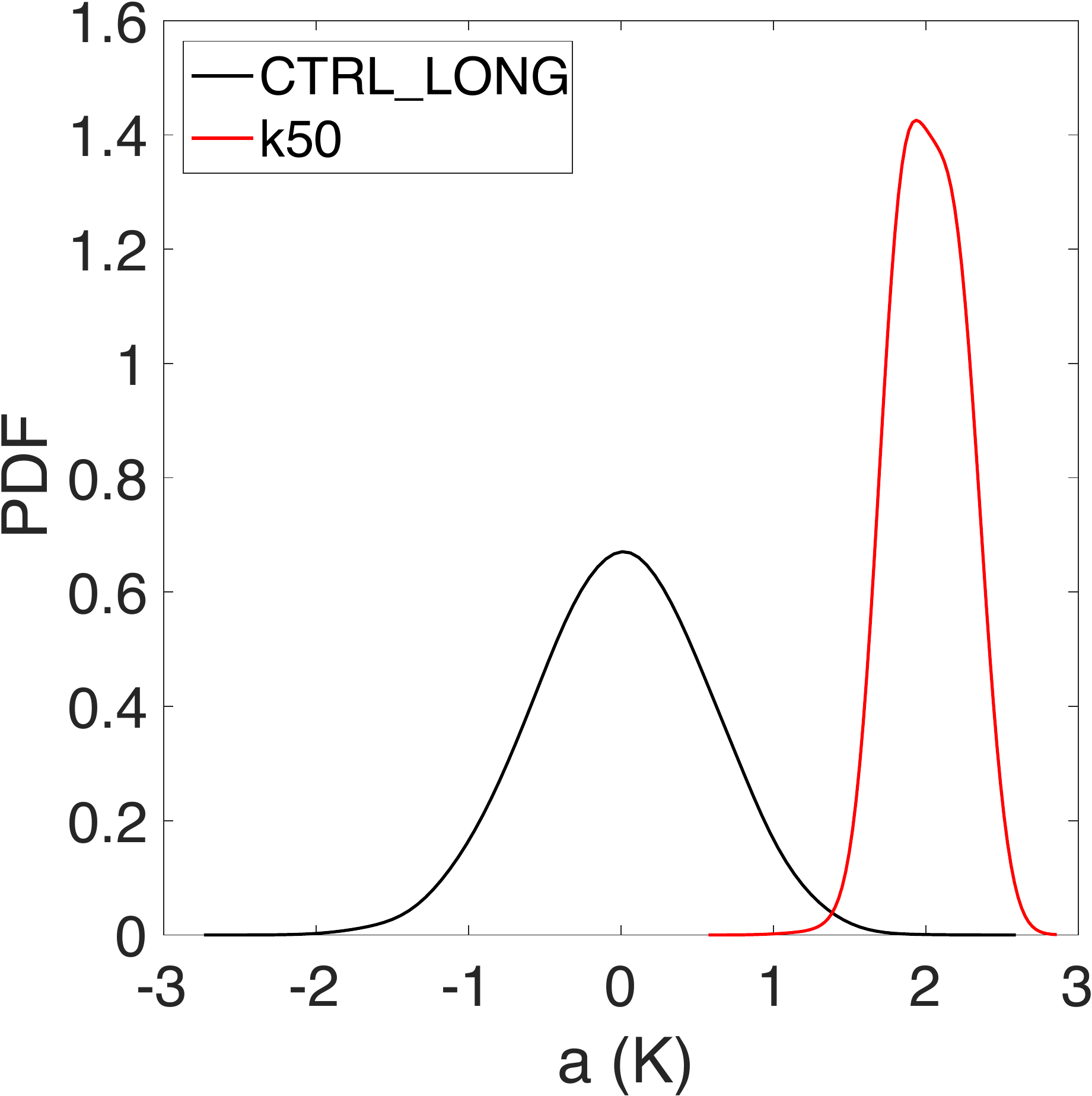}
\caption{(a) We want to estimate the probability to be in the set $B$, for
the model PDF $\rho(x)$. We are able to sample instead from the PDF
$\tilde{\rho}(x)$ for which the rare event becomes common. We know
the relation $L=\rho/\tilde{\rho}$ and can recover the model statistics
$\rho$, from the importance sampling $\tilde{\rho}$. (b) PDF of
the time average temperature $a$ ($T=90\,\text{d}$) for the model
control run (black) and for the algorithm statistics with $k=50$,
illustrating that the algorithm performs importance sampling and that $+2\text{\textdegree}\,K$
heat waves become common for the algorithm while they are rare for
the model.} 
\label{fig:Fig3}
\end{figure}

We first explain importance sampling, a crucial probabilistic concept
for the following discussion. We sample $N$ independent and identically
distributed random numbers from a probability distribution function
(PDF) $\rho$ and want to estimate $\gamma_{B}$, the probability
to be in a small set $B$ (see Fig. \ref{fig:Fig3}(a)). We will obtain
about $N\gamma_{B}$ occurrences in the set $B$, from which we can
estimate $\gamma_{B}$. An easy calculation \cite{rubino_rare_2009}
shows that the relative error of this estimate is of the order of
$1/\sqrt{N\gamma_{B}}$. For instance if $\gamma_{B}$ is of the order
of $10^{-2}$, estimating $\gamma_{B}$ with a relative error of $1\%$
requires a huge sample size, of the order of $10^{6}$. However, if
we rather sample $N$ random numbers $\tilde{X}_{n}$ from the distribution
$\tilde{\rho}$ (see Fig. \ref{fig:Fig3}(a)), where $\rho(x)=L(x)\tilde{\rho}(x)$,
with $L$ conveniently chosen, then the event may become common: this
is importance sampling. From the formula $\rho=L\tilde{\rho}$, we
have the estimate $\tilde{\gamma}_{B}=\frac{1}{N}\sum_{n=1}^{N}L\left(\tilde{X}_{n}\right)1_{B}\left(\tilde{X}_{n}\right)$,
where $1_{B}$ is the indicator function of the set $B$. If the rare
event is actually common for $\tilde{\rho}$, this estimate gives
a relative error of order $1/\sqrt{N}$ (see \cite{rubino_rare_2009}
for a precise formula). Then, in order to estimate $\gamma_{B}$ with
a relative error of $1\%$, we need a sample size of order of $10^{4}$;
this is a gain of a factor 100. The importance sampling gain grows
like the inverse of the probability $\gamma_{B}$. The key question
is: How to perform importance sampling, relevant for extreme heat
waves, starting from a climate model? 

Since the climate is a non equilibrium dynamical system, importance
sampling has to be performed at the level of the trajectories. Trajectories
generated by the model are distributed according to the unknown PDF
$\mathbb{P}_{0}\left(\left\{ X(t)\right\} _{0\leq t\leq T}=\left\{ x(t)\right\} _{0\leq t\leq T}\right)$
(this is a formal notation for the probability of the model variables
$X(t)$ to be close to $x(t)$). We use the GKLT large deviation algorithm, described bellow, 
that selects trajectories distributed according to the importance sampling
PDF $\mathbb{P}_{k}$
\begin{widetext}
\begin{align}
\mathbb{P}_{k}\left(\left\{ X(t)\right\} _{0\leq t\leq T}=\left\{ x(t)\right\} _{0\leq t\leq T}\right)=\frac{1}{Z(k,T)}\exp\left(k\int_{0}^{T}A\left(X(t)\right)\mbox{d}t\right)\mathbb{P}_{0}\left(\left\{ X(t)\right\} _{0\leq t\leq T}=\left\{ x(t)\right\} _{0\leq t\leq T}\right), \numberthis \label{eq:Tilted_Measure-1}
\end{align}
\end{widetext}
where $k$ is a real valued parameter, and $Z(k,T)$ is a normalization
constant such that $\mathbb{P}_{k}$ is a normalized PDF. The surface averaged temperature is $\tilde{A}(t)=A\left(X(t)\right)$. One observes
that for positive values of $k$, the measure $\mathbb{P}_{k}$ is tilted
with respect to $\mathbb{P}_{0}$ such that large values of $\int_{0}^{T}A\left(X(t)\right)\,\mbox{d}t$
will be favored with an exponential weight. Tuning $k$, we will study different ranges of extreme values for $a=\frac{1}{T}\int_{0}^{T}A\left(X(t)\right)\,\mbox{d}t$,
and thus different classes of extreme heat waves when $a$ is the
time averaged European temperature (\ref{eq:Observable-1}). 

The large deviation algorithm performs an ensemble simulation with $N$ trajectories (ensemble members), typically $N\sim O(10^2-10^3)$. The trajectories start from independent initial conditions that sample the model's invariant measure. After time intervals of constant duration $\tau$ we stop the simulation, and for each trajectory we  compute a score function based on the dynamics in the previous time interval of length $\tau$ (see the Supporting Information Appendix for the definition of the score function). Trajectories which are going in the direction of the extremes of interest, as measured by the score function, are cloned in one or more copies, while poorly scoring trajectories are killed. We call this step resampling, and $\tau$ the resampling time. The different copies of a successful trajectory are slightly perturbed, so that they can evolve differently. Then the ensemble of trajectories is iterated for another resampling time $\tau$. Once the final time $T_a$ has been reached, resampling is performed one last time. With a proper choice of the score function we obtain an ensemble of $N$ trajectories of length $T_a$ distributed according to equation (\ref{eq:Tilted_Measure-1}), where $k$ enters as a chosen parameter of the algorithm. The full details of the algorithm implementation are provided in the Supporting Information Appendix. 

In the normalization term of (\ref{eq:Tilted_Measure-1}), $Z(k,T)=\mathbb{E}_{0}\left[\mbox{e}^{k\int_{0}^{T}A\left(X(t)\right)\,\mbox{d}t}\right]$, the average is taken over the model statistics $\mathbb{P}_{0}$. In large deviation theory (see e.g. \cite{Touchette2009}), $\lambda(k)=\lim_{T\rightarrow\infty}\lambda(k,T)\,\,\,\mbox{with}\,\,\,\lambda(k,T)=\frac{1}{T}\log Z(k,t)$ is called a scaled cumulant generating function. One can prove that for large times, the PDF $P(a,T)$ of time averaged temperature $a$, satisfies $P(a,T)\underset{T\rightarrow\infty}{\asymp}\mbox{e}^{-TI\left[a\right]}$. Whenever $I$ is convex, $\lambda$ and $I$ are the Legendre\textendash Fenchel transform of one another: $\lambda(k)=\sup_{a}\left\{ ka-I(a)\right\} $ and $I(a)=\sup_{k}\left\{ ka-\lambda(k)\right\} $. The reader knowledgeable of statistical mechanics or thermodynamics will immediately notice the analogies between $Z$ and the partition function, $a$ and the energy, $k$ and the temperature, $\lambda$ and the free energy, and between $I$ and the entropy. To summarize, the large deviation algorithm allows us to choose the ``temperature'' $k$ for which dynamical states of ``energy'' $a$ (in this case time averaged European temperature) will become common. Increasing $k$ we can thus study events with more and more extreme heat waves.

\section*{Return times for 90 day heat waves}

\begin{figure}[h]
\centering
\includegraphics[width=0.6\linewidth]{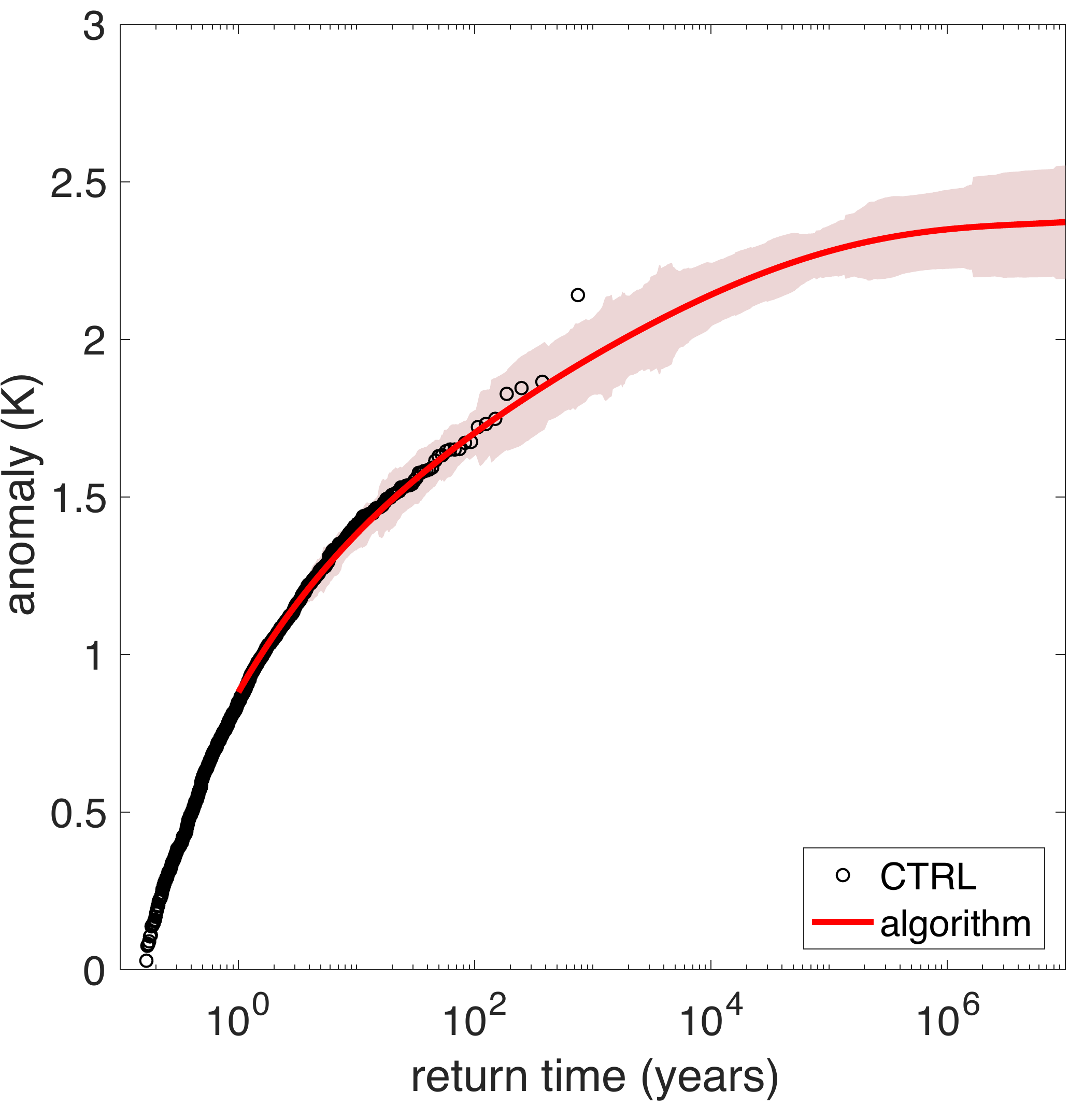}
\caption{Return times for the $90$ day Europe surface temperature, computed from
the $1000$ years long control run (black) and from the the large
deviation algorithm, at the same computational cost as the control
run (red). This illustrates both the good overlap on the $(10,300)$
year range, and the fact that the algorithm can predict probability
for events that can not be observed in the control run.}
\label{Fig4}
\end{figure}

We use the large deviation algorithm and formula (\ref{eq:Tilted_Measure-1})
in order to compute the return times for heat waves lasting several
weeks, following the methodology described in the Supporting Information Appendix.
Fig. \ref{Fig4} shows return times versus amplitude $a=\frac{1}{T}\int_{0}^{T}\tilde{A}(t)\,\mbox{d}t$,
for $T=90$ days. The black curve has been plotted from a $1\,000$
year control run. The red curve has been obtained as explained in
the Supporting Information Appendix from six experiments with the large deviation
algorithm with values of the bias parameter $k$ ranging from 10 to
40 (see Eq. (\ref{eq:Tilted_Measure-1})). Each of these simulations
has a computational cost of about $182$ years.

The first striking result on Fig. \ref{Fig4} is that we can
compute return times up to $10^{6}-10^{7}$ years with a total computational
cost of the order of $10^{3}$ years. This is thus a gain of more than
three orders of magnitude in the sampling efficiency. It is striking
that we can compute the return times for events that could not have
been observed in a direct numerical simulation with the current or
foreseeable computational possibilities. 

Another aspect is the improvement of the quality of the statistics. In the
control run there is only one heat wave with temperature in excess
of $2\,K$ during $90\,\text{days}$, while in the $k=50$ experiment
there are several hundreds, at a fraction of the computational cost.
We can thus recover the return time of such heat waves at either a
much smaller numerical cost compared with the control run, or with
a much smaller relative error, for a given numerical cost. Such an
improvement of the statistics will be crucial to perform a dynamical analysis 
that involves temperature and pressure fields, as discussed below.

\section*{Teleconnection patterns for extreme heat waves}

\begin{figure}[h]
\centering
(a)\includegraphics[width=0.6\linewidth]{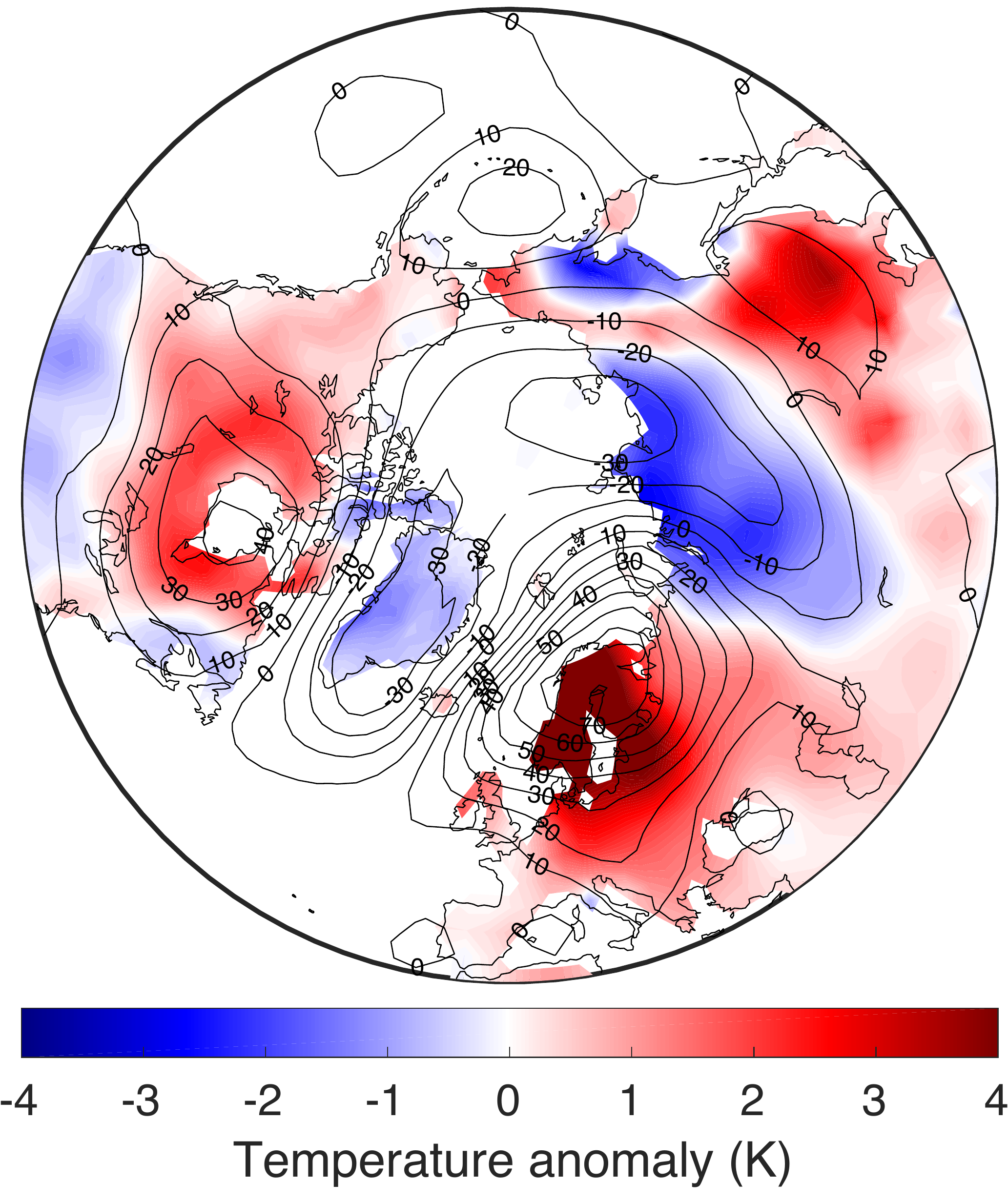}
(b)\includegraphics[width=0.6\linewidth]{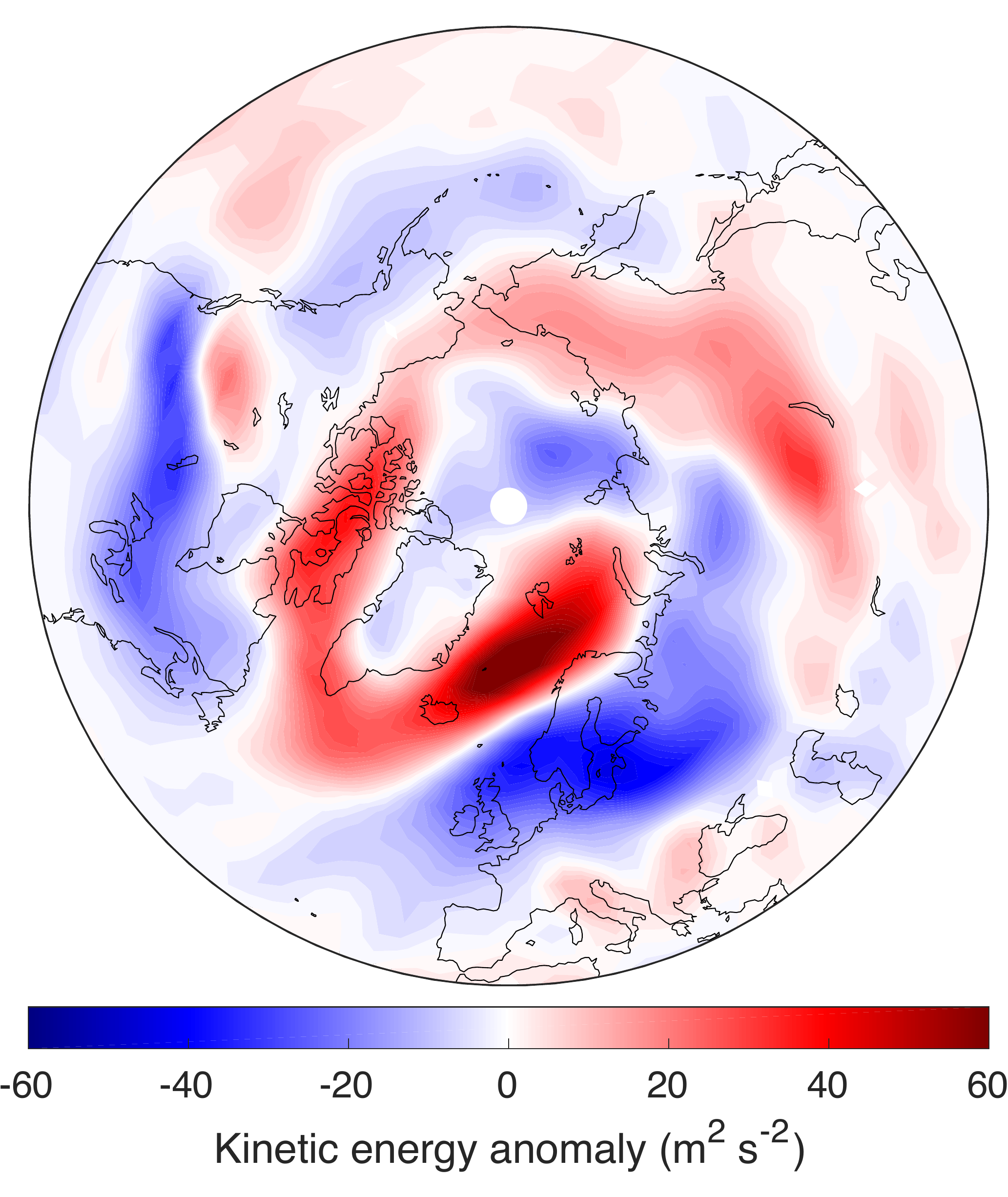}
\caption{(a) Northern hemisphere surface temperature anomaly (colors), and 500 hPa geopotential
height anomaly (contours), conditional on the occurrence of heat wave
conditions $\frac{1}{T}\int_{0}^{T}A(x_{n}(t))dt>a$, with $T=90\,\text{days}$
and $a=2\,K$, estimated from large deviation algorithm. (b) Northern hemisphere anomaly
of the averaged kinetic energy for the zonal velocity at 500 hPa conditional
on the occurrence of heat wave conditions $\mathbb{E}\left[KE_{500}\mid\frac{1}{T}\int_{0}^{T}A(x_{n}(t))dt>a\right]$,
with $T=90\,\text{days}$ and $a=2\,K$, estimated from the large
deviation algorithm, with respect to the long time average $\mathbb{E}\left[KE_{500}\right]$
computed from the control run.} 
\label{fig:Fig5}
\end{figure}
We use the excellent statistics gathered with the large deviation
algorithm in order to describe the corresponding state of the atmosphere
during extreme heat waves events. Fig. \ref{fig:Fig5}(a) shows the
temperature and the $500\,hPa$ geopotential height anomalies, conditioned
on the occurrence of a \textbf{$90$ }day $2\text{\textdegree\,K}$
heat wave (composite statistics). Those conditional statistics are
reminiscent of the teleconnection pattern maps sometimes shown in
the climate community. However while usual teleconnection patterns
are computed from empirical orthogonal function (EOF) analysis, and
thus describe typical fluctuations, our extreme event conditional
statistics describe very rare flows that characterize extreme heat
waves. Those global maps are a new way to consider rare event and
the atmosphere fluctuation statistics, which is extremely interesting
from a dynamical point of view. 

By definition, as we plot statistics conditioned on $a=\frac{1}{T}\int_{0}^{T}A(x_{n}(t))dt>2\text{\textdegree K}$,
with $T=90$ days, Fig. \ref{fig:Fig5}(a) shows a warming pattern
over Europe. The geopotential height map also shows a strong anticyclonic
anomaly right above the area experiencing the maximum warming, as
expected through the known positive correlation between surface temperature
and anticyclonic conditions \cite{Stefanon&al2012}. A less expected
and striking result is that the strong warming over Europe is correlated
with a warming over South Eastern Asia, and a warming over North America,
both with substantial surface temperature anomalies of order of $1\,K$
to $3\,K$, and anti-correlated with strong cooling over Russia and
Greenland, of the order of $-1\,K$ to $-2\,K$. This teleconnection
pattern is due to a strongly non linear stationary pattern for the
jet stream, with a wavenumber 3 dominating the pattern, as is clearly
seen from the geopotential height anomaly. On Fig. \ref{fig:Fig5}(b),
the anomaly of the kinetic energy, gives a complementary view: over
Europe, the succession of a southern blue band (negative anomaly),
and a northern red band (positive anomaly) should be interpreted as
a northward shift of the jet stream there. Strikingly, over Greenland
and North America, the jet stream is at the same position (but it
is more intense) for the large deviation algorithm statistics as for
the control run, while it is shifted northward over Europe and very
slightly southward over Asia. This is related to the strong southwest-northeast
tilt of the geopotential height anomalies over the Northern Atlantic.
The extended red
area (positive anomaly of kinetic energy) over Asia is rather due
to a more intense cyclonic activity there, than to a change of jet
stream position. 

Inspection of the time series of the daily temperature shows that
along the long duration of heat waves, the synoptic fluctuations on
times scales of weeks are still present (see also Figure \ref{fig:Fig2}b)).
The temperature is thus fluctuating with fluctuations of order of
5 to 10 degrees, as usual, but they fluctuate around a larger temperature
value than usual. This is also consistent with the northward shift
of the jet stream over Europe, but does not seem to be consistent
with a blocking phenomenology as hypothesized in many other publications.
This calls for using similar large deviation algorithms with other
models and other setups to test the robustness of the present observation. 

\section*{Conclusions}

We have demonstrated that rare event algorithms, developed using statistical
physics ideas, can improve the computation of the return times and
the dynamical aspects of extreme heat waves. One of the future challenge in the use of rare event algorithms for studying climate extreme will be to identify which algorithms and which score functions will be suitable for each type of rare event. We anticipate that this
new tool will open a range of completely new studies that were out
of scope so far. First it will pave the way to the use of state of
art climate model to study rare extreme events, without having to
run the model for unaffordable times. The demonstrated gain of several\textbf{
}orders of magnitude in the sampling efficiency will also help to
make quantitative model comparison, in order to assess on a more quantitative
basis the skill to predict extreme events, for the existing models. 
It will also open a new range of dynamical studies. As an example,
having an high number of heat waves allowed us to conclude that a
Europe heat wave, mainly affecting Scandinavia, is rather related
to a Northward jet-stream shift rather than a Rossby wave breaking,
in the Plasim model. Such a phenomenology may well be model and model
resolution dependent. Finally, and may be more importantly, this new
tool will be extremely useful in the near future to assess quantitatively
anthropogenic carbon dioxide emission impact on heat waves and other
classes of extreme events. Assessment of the anthropogenic causes of rare event return time changes requires to compare two different climates \cite{intergovernmentalpanelonclimatechange_climate_2013,Otto&al2012}, and to run a rare event algorithm for each case.

\section*{SI Appendix}

The Supporting Information Appendix contains a complete description of the GKLT algorithm, of the method to compte return times with rare event algorithm, the description of the implementation of the Plasim model, aspects of the statistical post processing, and the description of the dynamical quantities represented in this article.



\acknow{The authors would like to thank Gualtiero Badin, Edilbert Kirk, and
Thibault Lestang for useful discussions and suggestions on various
aspects of this work. JW and FR acknowledge the support of the AXA
Research Fund.}

\showacknow 


\bibliography{PNAS_Heatwaves}

\clearpage

\section*{Supporting Information (SI) appendix}

\section*{SI Data and Methods}

\subsection*{Giardina\textendash Kurchan\textendash Tailleur\textendash Lecomte
algorithm}

Rare event algorithms have first been proposed and developed in the
50's \cite{kahn_estimation_1951}, and they have since then been used
for a wide range of applications. A partial mathematical analysis
of these algorithms is now available \cite{delmoral_feynmankac_2004}. 

The general idea is the following. We let evolve in parallel an ensemble
of trajectories of a numerical model starting from different initial
conditions. After a given \textit{resampling time} some members of
the ensemble are killed and some others are cloned, depending on the
values of weights defined on the past evolution of the trajectories.
In this way the trajectory probability measure is tilted with respect
to the ``natural'' model trajectory probability measure. How the
distribution will be tilted will depend on the definition of the weights,
which can be defined in order to target the extremes of an observable
of interest.

In the algorithm used in this work, the weights are chosen such that
the measure is tilted in order to favor trajectories characterized
by large values of a chosen time averaged observable. This form of
a rare event algorithm has first been proposed by \cite{giardina_direct_2006}\textbf{
}and has been used for instance to study finite time Lyapunov exponents
\cite{tailleur_probing_2007}. As the primary aim of this algorithm
is to compute large time large deviation rate function, we call this
algorithm the \textit{Giardina-Kurchan-Lecomte-Tailleur large deviation algorithm}, or simply the \textit{large deviation
algorithm}.

We perform simulations of an ensemble of $N$ trajectories $\left\{ X_{n}(t)\right\} $
(with $n=1,2,...,N$) starting from different initial conditions.
The total integration time of the trajectories is denoted $T_{a}$.
In the limit of large $T_{a}$, the initial conditions affect only
a transient regime. We consider an observable of interest $A(X(t))$
(in this study the Europe temperature) and a resampling time $\tau$.
At times $t_{i}=i\tau$ (with $i=1,2,...,T_{a}/\tau$) we assign to
each trajectory $n$ a weight $W_{n}^{i}$ defined as 
\begin{equation}
W_{n}^{i}=\frac{e^{k\intop_{t_{i-1}}^{t_{i}}A(X_{n}(t))dt}}{R_{i}}\,\,\,\mbox{with}\,\,\,R_{i}=\frac{1}{N}\sum_{n=1}^{N}e^{k\int_{t_{i-1}}^{t_{i}}A(X_{n}(t))dt}.\label{eq:Weight}
\end{equation}
For each trajectory $X_{n}$, a random number of copies of the trajectory
are generated, on average proportional to the weight $W_{n}^{i}$
and such that the total number of trajectories produced at each event
is equal to $N$. The parameter $k$ is chosen by the user in order
to control the strength of the selection and thus to target a class
of extreme events of interest. The larger the value of $k$, the more
trajectories with large values of the time average observable will
survive the selection. 

Let us denote $\mathbb{P}_{0}\left(\left\{ X(t)\right\} _{0\leq t\leq T_{a}}\right)$
the probability to observe a trajectory $\left\{ X(t)\right\} _{0\leq t\leq T_{a}}$
in the model, and $\mathbb{P}_{k}\left(\left\{ X(t)\right\} _{0\leq t\leq T_{a}}\right)$
the probability to observe the same trajectory with the algorithm.
By construction of the algorithm through the weights (\ref{eq:Weight})
we have that
\begin{widetext}
\begin{align}
\mathbb{P}_{k}\left(\left\{ X(t)\right\} _{0\leq t\leq T_{a}}=\left\{ x(t)\right\} _{0\leq t\leq T_{a}}\right)\underset{N\rightarrow\infty}{\sim}\frac{e^{k\int_{0}^{T_{a}}A(x(t))dt}}{\mathbb{E}_{0}\left[e^{k\int_{0}^{T_{a}}A(x(t))dt}\right]}\mathbb{\mathbb{P}}_{0}\left(\left\{ X(t)\right\} _{0\leq t\leq T_{a}}=\left\{ x(t)\right\} _{0\leq t\leq T_{a}}\right).\label{eq:Biased_Path_Approximation}
\end{align}
\end{widetext}
where $\mathbb{E}_{0}$ mean an average over $\mathbb{P}_{0}$, and
$\underset{N\rightarrow\infty}{\sim}$ means that this is true only
asymptotically for large $N$ with typical error of order $1/\sqrt{N}$
when evaluating averages over observables. Equation (\ref{eq:Biased_Path_Approximation})
is obtained by assuming the mean field approximation 
\begin{equation}
R_{1}=\frac{1}{N}\sum_{n=1}^{N}e^{k\int_{0}^{t_{_{1}}}A(X_{n}(t))dt}\underset{N\rightarrow\infty}{\sim}\mathbb{E}_{0}\left[e^{k\int_{0}^{t_{1}}A(X_{n}(t))dt}\right],\label{eq:Mean_Field_Approximation}
\end{equation}
which, by induction, and using a formula similar to (\ref{eq:Mean_Field_Approximation})
at each step of the induction, leads to 
\begin{equation}
\prod_{i=1}^{T_{a}/\tau}R_{i}\underset{N\rightarrow\infty}{\sim}\mathbb{E}_{0}\left[e^{k\int_{0}^{t_{1}}A(X_{n}(t))dt}\right]\label{eq:Estimate_Lambda}
\end{equation}
(see \cite{giardina_direct_2006,giardina_simulating_2011}).

The validity of the mean field approximation and the fact that the
typical relative error due to this approximation is of order $1/\sqrt{N}$
has been proven to be true for a family of rare event algorithms including
the one adopted in this paper by \cite{delmoral_feynmankac_2004}.

Formula (\ref{eq:Biased_Path_Approximation}) is valid only for times
$T_{a}$ that are integer multiples of the resampling time $\tau$.
The killed trajectories have to be discarded from the statistics.
Starting from the final $N$ trajectories at time $T_{a}$, one goes
backwards in time through the selection events attaching to each piece
of trajectory its ancestor. In this way one obtains an effective ensemble
of $N$ trajectories from time 0 to time $T_{a}$, distributed according
to $\mathbb{P}_{k}$. All trajectories reconstructed in this way are
real solutions of the model, so that we are not changing the dynamics,
but only sampling trajectories according to the distribution $\mathbb{P}_{k}$
rather than according to the distribution $\mathbb{P}_{0}$.

In the normalization term $Z(k,T)=\mathbb{E}_{0}\left[\mbox{e}^{k\int_{0}^{T}A\left(X(t)\right)\,\mbox{d}t}\right]$, the average is taken over the model statistics $\mathbb{P}_{0}$. In large deviation theory (see e.g. \cite{Touchette2009}), $\lambda(k)=\lim_{T\rightarrow\infty}\lambda(k,T)\,\,\,\mbox{with}\,\,\,\lambda(k,T)=\frac{1}{T}\log Z(k,t)$ is called a scaled cumulant generating function. One can prove that for large times, the PDF $P(a,T)$ of time averaged temperature $a$, satisfies $P(a,T)\underset{T\rightarrow\infty}{\asymp}\mbox{e}^{-TI\left[a\right]}$. Whenever $I$ is convex, $\lambda$ and $I$ are the Legendre\textendash Fenchel transform of one another: $\lambda(k)=\sup_{a}\left\{ ka-I(a)\right\} $ and $I(a)=\sup_{k}\left\{ ka-\lambda(k)\right\} $. The reader knowledgeable of statistical mechanics or thermodynamics will immediately notice the analogies between $Z$ and the partition function, $a$ and the energy, $k$ and the temperature, $\lambda$ and the free energy, and between $I$ and the entropy. To summarize, the large deviation algorithm allows us to choose the ``temperature'' $k$ for which dynamical states of ``energy'' $a$ (in this case time averaged European temperature) will become common. 

By construction \cite{giardina_direct_2006,giardina_simulating_2011}
the algorithm gives an estimator of the scaled cumulant generating
function of $A(X(t))$ 
\begin{equation}
\lambda(k)=\lim_{T\rightarrow\infty}\lambda(k,T)\,\,\,\text{with}\,\,\,\lambda(k,T_{a})=\frac{1}{T_{a}}\sum_{i=1}^{T_{a}/\tau}\log R_{i},\label{eq:GK_lambda_estimator}
\end{equation}
with a relative error of order $1/\sqrt{N}$. 

While the GKLT algorithm has been initially designed to compute large
deviation rate function, we can use it to compute any statistical
quantity related to the statistic $\mathbb{\mathbb{P}}_{0}$, from
the $\mathbb{P}_{k}$. This is done using the backward reconstructed
trajectories and inverting formula (\ref{eq:Biased_Path_Approximation}).
If one for example want to estimate the expectation value of an observable
$O\left(\left\{ X(t)\right\} _{0\leq t\leq T_{a}}\right)$ of the
model, an estimator is thus
\begin{widetext}
\begin{align}
\mathbb{E}_{0}\left[O\left(\left\{ X(t)\right\} _{0\leq t\leq T_{a}}\right)\right]\underset{N\rightarrow\infty}{\sim}\frac{1}{N}\sum_{n=1}^{N}O\left(\left\{ X_{n}(t)\right\} _{0\leq t\leq T_{a}}\right)\mbox{e}^{-k\int_{0}^{T_{a}}A(X_{n}(t))dt}\mbox{e}^{T_{a}\lambda(k,T_{a})},\label{eq:GK_O_estimator}
\end{align}
\end{widetext}
where the $X_{n}$ are the $N$ backward reconstructed trajectories.
Empirical estimators of quantities related to rare (for $\mathbb{\mathbb{P}}_{0}$)
events of the kind of \ref{eq:GK_O_estimator} (thus using data distributed
according to $\mathbb{P}_{k}$) have a dramatically lower statistical
error, due to the larger number of relevant rare events present in
the effective ensemble.

\subsection*{Algorithm implementation}

We describe the large deviation algorithm implementation. Start with
$N$ trajectories, each with different initial conditions, for instance
sampled through an ensemble of statically independent states obtained
from a control run. For $i=1,..,T_{a}/\tau$ 
\begin{enumerate}
\item Iterate each trajectory from time $t_{i-1}=(i-1)\tau$ to time $t_{i}=i\tau$; 
\item At time $t_{i}$ stop the simulation and assign to each trajectory
$n$ a weight 
\begin{equation}
W_{n}^{i}=\frac{e^{k\intop_{t_{i-1}}^{t_{i}}A(X_{n}(t))\,\mbox{d}t}}{R_{i}}
\end{equation}
with
\begin{equation}
R_{i}=\frac{1}{N}\sum_{n=1}^{N}e^{k\int_{t_{i-1}}^{t_{i}}A(X_{n}(t))\,\mbox{d}t}.
\end{equation}
\item Compute the number of copies produced by each trajectory as 
\begin{equation}
c_{n}^{i}=\left\lfloor W_{n}^{i}+u_{n}\right\rfloor 
\end{equation}
where $\left\lfloor \cdot\right\rfloor $ is the integer part and
the $u_{n}$ are $N$ independent random numbers uniformly distributed
between 0 and 1. When $c_{n}^{i}=0$ the corresponding trajectory
is killed; 
\item The number of trajectories present after the selection operation is
\begin{equation}
N_{i}=\sum_{n=1}^{N}c_{n}^{i}
\end{equation}
 
\item Compute then the difference $\Delta N_{i}=N_{i}-N$. If $\Delta N_{i}>0$,
then $\Delta N_{i}$ trajectories are randomly selected (without repetition)
and killed. If $\Delta N_{i}<0$, then $\Delta N_{i}$ trajectories
are randomly selected (with repetition) and cloned. 
\item Reinitialize the state of the $N$ ensemble members at time $t_{i}$
according to the cloning, and restart from point 1 incrementing $i$
by 1. 
\end{enumerate}
The $\Delta N_{i}$ trajectories to be killed/cloned are chosen among
the $N_{i}$ trajectories present after the killing/cloning event,
not among the $N$ trajectories present before the killing/cloning
event. In the case $\Delta N_{i}<0$, only trajectories for which
$c_{n}^{i}>0$ are allowed to be cloned. Operations 3-5 guarantee
that the number of ensemble members remains constant throughout the
evolution of the system \cite{giardina_simulating_2011}. After an
initial transient the algorithm ensemble reach a statistically stationary
state, and the memory of the initial conditions is lost.

The $N$ initial conditions should be independent and provide a reasonable sampling of the attractor of the system.
A possible way to create the initial conditions is to generate a control run of the model, and take states of the system
separated by long enough times so that they are statistically independent. In our study we have created a 1000 year control run (also used as a benchmark for the performances of the algorithm). From this, we have taken 1000 model states, at one year intervals, as our set of initial conditions. If we would have not needed a control run, we could have used initial conditions taken every week (which is about the correlation time of the dynamics for this model). It is essential to note that the duration of the run to prepare independent initial conditions is smaller by several orders of magnitudes than the return times that can be computed using the algorithm. 

Tests performed with the Ornstein-Uhlenbeck process and the Lorenz
63 model suggest that the resampling time $\tau$ should be not larger
than the Lyapunov time of the model, and sufficiently larger than
the numerical time step. Between these two limits the performances
of the algorithm seems to be insensitive to the specific choice of
$\tau$. Here we have taken $\tau=8$ days, close to the autocorrelation
time of the European surface temperature in Plasim, that is $\tau_{c}\approx7.5$
days.

Dealing with a deterministic system, two clones of the same trajectory
will evolve exactly in the same way. In order to have copies of the
same trajectory that actually separate in the following time evolution,
we add a small random perturbation to the state of the trajectories
at time $t_{i}$, immediately after the reinitialization according
to the cloning \cite{wouters_rare_2015}. The perturbation is introduced
adding to the coefficients of the spherical harmonics of the logarithm
of the surface pressure a set of random numbers, sampled independently
according to a uniform distribution in $[-\epsilon\sqrt{2},\epsilon\sqrt{2}]$,
with $\epsilon=10^{-4}$. On average the relative perturbation (computed
as the difference between the root mean square of the spherical harmonic
coefficients before and after the perturbation, divided by the root
mean square of the spherical harmonic coefficients) is of order $\sim7.5\cdot10^{-5}$.
Assuming linear response to external perturbations for the statistics
of the system (the invariant measure of the system) \cite{Ruelle1998,Ruelle2009},
under the the Chaotic Hypothesis \cite{Gallavotti1996} (exploiting
the chaoticity of the dynamics and the large number of degrees of
freedom), the statistical properties of the system are thus expected
to be altered by an error of order of the perturbation. This is way
lower than any other errors, for instance sampling errors.

\subsection*{Plasim model and setup}

The numerical model used in this study is the Planet Simulator (Plasim)
\cite{Fraedrich&al2005}. Plasim is a spectral intermediate
complexity general circulation model with a full set of physical parameterizations
for unresolved processes. It produces a fairly realistic present climate
and is representative of the class of complex numerical models used
for climate prediction, although it is computationally less demanding
than contemporary IPCC-standard climate models. See the Reference
Manual at \url{https://github.com/Edilbert/PLASIM} for a detailed
description.

We set the model at T42 horizontal resolution and 10 levels vertical
resolution, for a total of $O(10^{5})$ degrees of freedom. The time
step is 30 minutes and output is stored every 6 hours. The model runs
in perpetual summer conditions, with no daily nor seasonal cycle.
The boundary conditions (sea surface temperature, ice coverage, and
radiative forcing at top of the atmosphere) are set to climatological
values for July 16th. The choice of a perpetual summer simulation
is for convenience: running the large deviation algorithm is simpler
in this context. We will add a seasonal cycle in a forthcoming work.
We consider as test observable the surface temperature in the atmospheric
layer $2\mathrm{m}$ above the ground, averaged over the land area
with latitudinal and longitudinal boundaries 36N-70N, 11W-25E respectively
(so called Europe surface temperature in the main text).

A 1000 year long control run (CTRL\_LONG) is taken as a benchmark
for the statistics. A shorter run (CTRL\_SHORT) consisting of the
last 284 years of the CTRL\_LONG run is used to compare the performances
of the algorithm against direct sampling.

The experiment with the large deviation algorithm with $k=2$ is performed
with $N=128$ trajectories. The integration time of each trajectory
is $T_{a}=800$ days, thus the computational cost is $C=NT_{a}\approx284.4$
years. The resampling time is set at $\tau=8$ days, which is of the
order the autocorrelation time $\tau_{c}=7.5$ days of the Europe
averaged temperature. 

\renewcommand{\thefigure}{S\arabic{figure}}

\begin{figure}[h] 
\centering 
\includegraphics[width=0.6\linewidth]{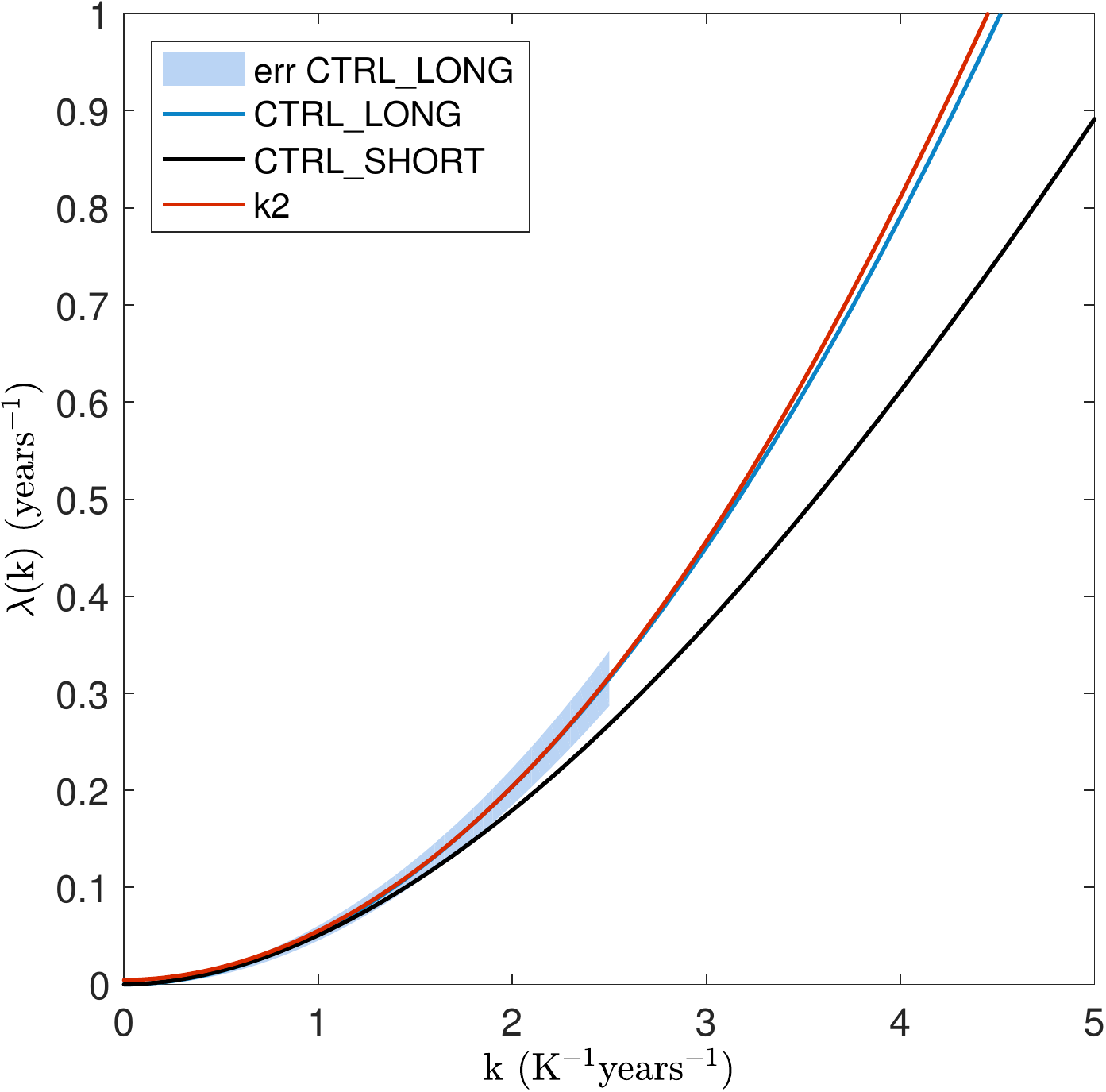} 
\caption{The scaled cumulant generating function $\lambda(k)$ estimated from the large deviation algorithm with $k=2$ at a numerical cost of 248 years (red line), is at least as good as the one estimated from a control run lasting 1000 years (blue line), and is much better than the one estimated from a control run lasting 248 years (black line). The shaded area corresponds to 1 standard deviation related to the empirical average.} 
\label{fig:FigS1} 
\end{figure}

Figure \ref{fig:FigS1} shows how the scaled cumulant generating function $\lambda$ estimated from the large deviation algorithm with $k=2$ at a numerical cost of 248 years, is at least as good as the one estimated from a control run lasting 1000 years, and is much better than the one estimated from a control run lasting 248 years. This is a consistency test between the large deviation algorithm and the direct use of the model, which confirms the robustness of the procedure we have followed (including the perturbation of the trajectories).

We have then performed six experiments with much larger values of
$k$, in order to compute efficiently the statistics of extreme heat
waves. We have taken $k=10,20,40$ and $50$. The experiments with
$k=20$ and $40$ have been performed twice with different initial
conditions. We have taken $N=512$ trajectories, each $128$ days
long, and the resampling time is again set to 8 days.

In order to demonstrate that the algorithm actually reduces the variance, we have computed the relative error on the estimate of surface temperature and of the 500 hPa geopotential height, comparing a control run and the statistics of the large deviation algorithm. The results (not presented) show that the empirical variance is way lower using the large deviation algorithm, than using the control run.

\subsection*{Plotting return time curves using the large deviation algorithm\label{subsec:Plotting-return-time}}

\subsubsection*{Return time plots for rare events sampled from a timeseries}

Let us consider a stochastic process $\left\{ Y(t)\right\} $ and
a threshold value $a$. We define the random variable $\tau(a,t)=\min\left\{ \tau\geq t\left|Y\left(\tau\right)>a\right.\right\} $.
Then the return time with threshold $a$ is defined as 
\begin{equation}
r(a)=\mathbb{E}\left(\tau(a,t)\right),\label{eq:Return_Times_Definition}
\end{equation}
where $\mathbb{E}$ is the average over the realization of the process.
Let us consider the estimate of $r(a)$ from a sample timeseries of
duration $T_{d}$: $\left\{ Y(t)\right\} _{0\leq t\leq T_{d}}$. Let
us consider a rare value $a$ such that most of the times $Y(t)<a$,
or equivalently $r(a)\gg\tau_{c}$ where $\tau_{c}$ is the correlation
time of the process. The return time $r(a)$ is then the average time
one has to wait in between two statistically independent events exceeding
the value $a$.

Let us divide the timeseries $\left\{ Y(t)\right\} _{0\leq t\leq+\infty}$
in pieces of duration $\Delta T\gg\tau_{c}$, and let us define\\
$a_{m}=\max\left\{ Y(t)\left|(m-1)\Delta T\leq t\leq m\Delta T\right.\right\} $
and $s_{m}(a)=1$ if $a_{m}>a$ and $0$ otherwise. With generic hypothesis
on the loss of memory for the process, the number of events $N(t)=\sum_{m}s_{\left\lfloor t/\Delta T\right\rfloor }$
is then well approximated by a Poisson process with density $\lambda(a)=1/r(a)$,
asymptotically when $r(a)\gg\tau_{c}$. An estimate of $r(a)$ is
then
\begin{equation}
r(a)=-\frac{\Delta T}{\log\left(1-\frac{1}{M}\sum_{m=1}^{M}s_{m}(a)\right)}.\label{eq:Return_Times_Rare-1-1}
\end{equation}

We then rank the sequence $\left\{ a_{m}\right\} _{1\leq m\leq M}$
in decreasing order and denote the ranked sequence $\left\{ \tilde{a}_{m}\right\} _{1\leq m\leq M}$
such that $\tilde{a}_{1}\geq\tilde{a}_{2}\geq...\geq\tilde{a}_{M}$.
Using the estimate (\ref{eq:Return_Times_Rare-1-1}) we associate
to the threshold $\tilde{a}_{m}$ the return time $r\left(\tilde{a}_{m}\right)=-\frac{\Delta T}{\log\left(1-\frac{m}{M}\right)}$.
The return time plot is then the plot $r\left(\tilde{a}_{m}\right)$
as a function of $\tilde{a}_{m}$, as illustrated for instance on
figure \ref{fig:FigS2}. 

\subsubsection*{Return time plots sampled using the large deviation algorithm}

The large deviation algorithm provides an effective ensemble of trajectories.
For each of these trajectories, we compute $A_{n}(T,t)=\frac{1}{T}\int_{t}^{T+t}A\left(X_{n}(t)\right)\,\text{dt}$
and compute maxima over the trajectory $a_{n}=\max_{0\leq t\leq T_{a}-T}\left\{ A_{n}(T,t)\right\} $.
Rather than providing just a sequence $\left\{ a_{n}\right\} _{1\leq n\leq N}$,
the large deviation algorithm provides a sequence $\left\{ a_{n},p_{n}\right\} _{1\leq n\leq N}$
where each trajectory, and thus each maxima $a_{n}$, is associated
with a probability $p_{n}$ of observing this value in the model.
The probability $p_{n}$ is computed from equation (\ref{eq:GK_O_estimator}):
$p_{n}=\frac{1}{N}\mbox{e}^{-k\int_{0}^{T_{a}}A(X_{n}(t))dt}\mbox{e}^{T_{a}\lambda(k,T_{a})}$.
As a consequence, the generalization of formula (\ref{eq:Return_Times_Rare-1-1})
is straightforward and leads to 

\begin{equation}
r(a)=-\frac{T_{a}-T}{\log\left(1-\sum_{n=1}^{N}p_{n}s_{n}(a)\right)},\label{eq:Return_Time_Large_Deviation_Algorithm-1}
\end{equation}
(we recall that $s_{n}(a)=1$ if $a_{n}>a$ and $0$ otherwise, $\Delta T=T_{a}-T$
is the length of the considered timeseries for each $A_{n}(T,t)$).
In practice, to plot the return time curve, we rank the sequence $\left\{ a_{n},p_{n}\right\} _{1\leq n\leq M}$,
sort it in decreasing order for the values of $a$, and denote the
ranked sequence $\left\{ \tilde{a}_{n},\tilde{p}_{n}\right\} _{1\leq n\leq M}$
such that $\tilde{a}_{1}\geq\tilde{a}_{2}\geq...\geq\tilde{a}_{N}$.
We then associate to the threshold $\tilde{a}_{m}$ the return time
\begin{equation}
r(\tilde{a}_{n})=-\frac{T_{a}-T}{\log\left(1-\sum_{m=n}^{N}\tilde{p}_{m}\right)},\label{eq:Return_Time_Large_Deviation_Algorithm}
\end{equation}
as the average number of events that have been observed with an amplitude
larger than $\tilde{a}_{n}$ is $\sum_{m=n}^{N}\tilde{p}_{m}$. The
return time plot is then the plot $r\left(a\right)$ as a function
of $a$, as illustrated for instance on figure \ref{fig:FigS2}. 

As a benchmark, the following section illustrates that return time
plots sampled from the algorithm coincide the return time sampled
from the timeseries of the Ornstein\textendash Uhlenbeck process and
can be computed at a much smaller computation cost.

\subsubsection*{Test of the algorithm for the Ornstein\textendash Uhlenbeck process}

\begin{figure}[h]
\centering
(a)\includegraphics[width=0.6\linewidth]{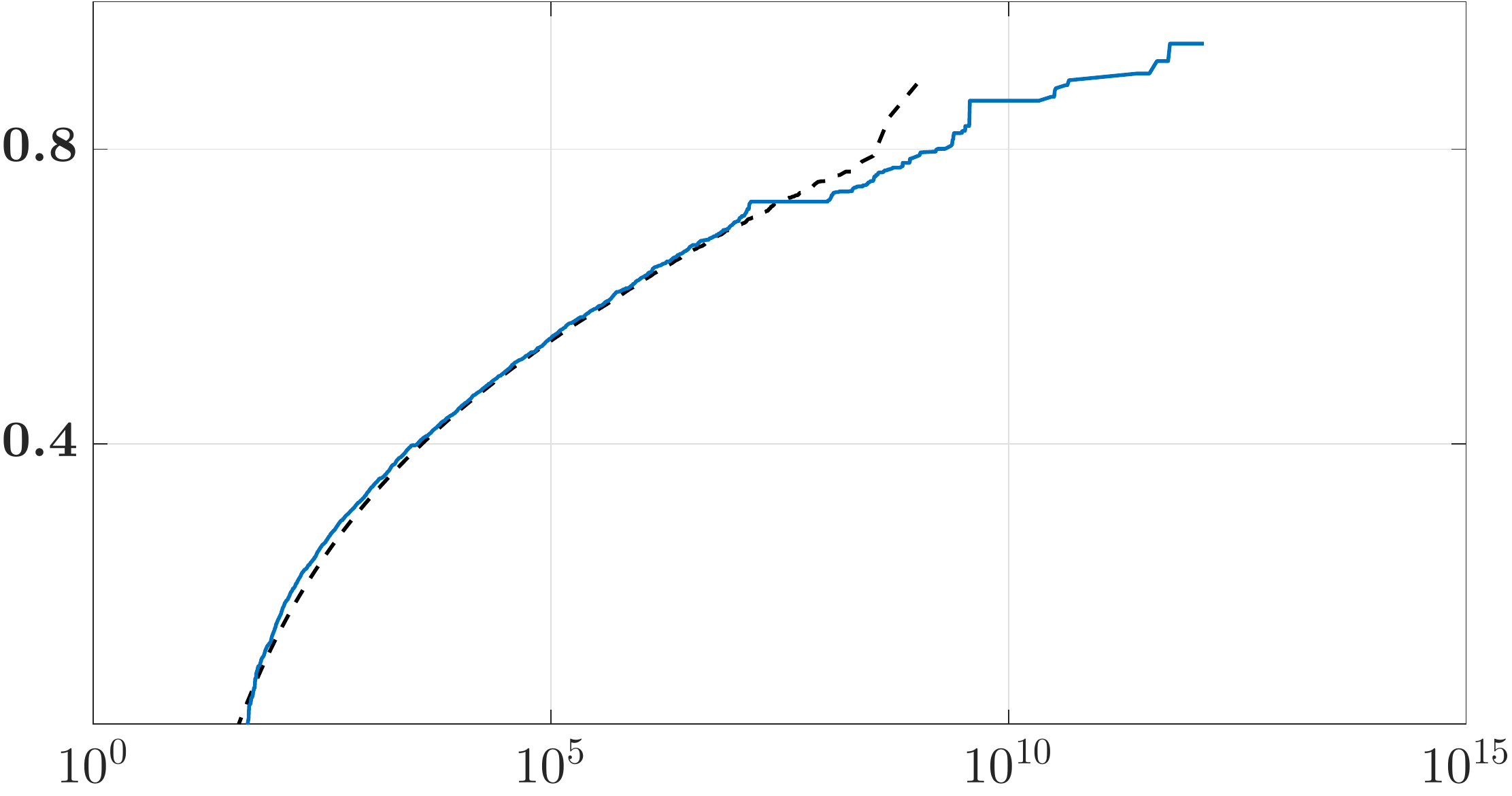}
(b)\includegraphics[width=0.6\linewidth]{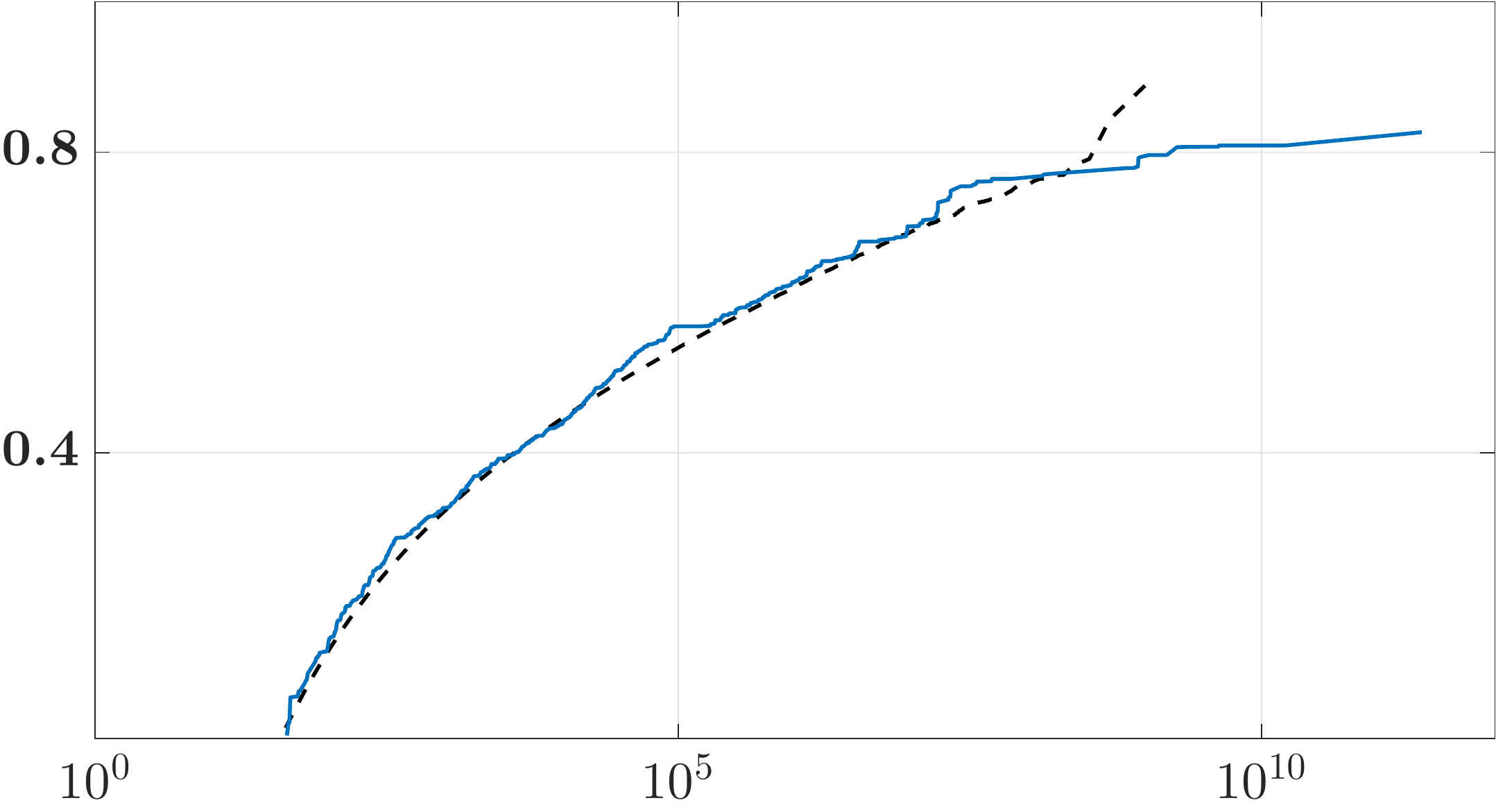}
\caption{Amplitude versus return times for the Ornstein--Uhlenbeck process, from a long control run (black dashed line) and the large deviation algorithm (blue curve). On panel a) 100 algorithm realizations have been used for a total cost $160$ smaller than for the control run. On panel b) 10 algorithm realizations have been used for a total cost $1600$ smaller than for the control run.} 
\label{fig:FigS2}
\end{figure}

Figure \ref{fig:FigS2} shows the algorithm benchmark for the
Ornstein\textendash Uhlenbeck process: the dynamics $\text{dx=-x\ensuremath{\text{d}t}+d}W_{t}$
where $W_{t}$ is the Wiener process ($\frac{\text{d}W_{t}}{\text{d}t}$
is a Gaussian noise with correlations $\mathbb{E}\left(\frac{\text{d}W_{t}}{\text{d}t}\frac{\text{d}W_{t'}}{\text{d}t}\right)=\delta\left(t-t'\right)$).
We study rare events for the time average observable $A(t)=\frac{1}{T}\int_{t}^{T+t}x(t)\,\text{d}t$
with $T=50$. $r(a)$ is the return time for this observable with
amplitude $a$ (please see definition (\ref{eq:Return_Times_Definition})).
We first sample $r(a)$, using formula (\ref{eq:Return_Times_Rare-1-1}),
from a single realization of the Ornstein\textendash Uhlenbeck process,
of total duration $T_{d}=10^{9}$ (control run). The result is shown
as a black dashed line on both figures \ref{fig:FigS2}a) and
b). The blue plain curve on figure \ref{fig:FigS2}a) has been
sampled using the large deviation algorithm, with a duration per trajectory
$T_{a}=100$, a clone number $N=600$, and a cloning time $\tau=0.5$.
The algorithm has been run $K=100$ times to gather statistics. The
total numerical cost is thus $T_{tot}=T_{a}*N*K=6.10^{6}$, which
is $160$ times less than the total duration $T_{d}$ for the control run. The
parameter $k$ has been set to $k=0.3$ such that events with a threshold
$0.6<a<0.7$ are typical. The result shows a perfect comparison with
the control run up to return times of order $5.10^{7}$ (much longer
than the total numerical cost) and an agreement with error of order
$10\%$ for return times $5.10^{7}<r(a)<5.10^{8}$. This shows that
the algorithm computes extremely well the return time plot at a much
lower numerical cost.  Figure \ref{fig:FigS2}a) show the same results, but using only $10$ algorithm realizations for a total cost  $1600$ times less than the total duration $T_{d}$ for the control run.

\subsection*{Statistical postprocessing}

All the statistical postprocessing has been performed in Matlab environment.
In particular, the probability distribution functions in Fig. 3(b)
have been computed using a kernel density estimation method.

The direct estimate of the scaled cumulant generating function from
the control run has been obtained following \cite{PhysRevE.92.052104}.
In particular, the error bar in Fig. 4(a) is computed taking one standard
deviation of the empirical sum of the estimator proposed in \cite{PhysRevE.92.052104},
limited to the region of Gaussian convergence.

The estimate of the return time curve in general has been obtained
as explained in Section 4. In particular, the estimate from the algorithm
(red line of Fig 4(b)) has been obtained by joining the results of
six experiments, as follows. First we have computed for each experiment
the return time curve $r(a)$ using formula (\ref{eq:Return_Time_Large_Deviation_Algorithm}),
which gives plots $a(r)$. We have linearly interpolated the plots
$a(r)$ on an equally spaced return time vector $R$ ranging from
$0$ to $10^{8}$ years. We have removed from each interpolated return
time curve the anomalies that are outside one standard deviation around
the mean, as we want to keep only the estimates that we consider reliable.
We have then obtained a best estimate return time curve by averaging
the tapered and interpolated return time curves from the individual
experiments. Finally, the best estimate curve has been interpolated
with a fifth order polynomial. The error bar is computed at each value
of $R$ as one standard deviation of the return time curves which
have been averaged in order to obtain the best estimate at that value
of $R$ (usually 3 or 4 curves).

\subsection*{Featured dynamical quantities}

The \textit{geopotential} $\Phi(\phi,\lambda,z,t)$ is the gravitational
potential energy per unit mass at latitude $\phi$, longitude $\lambda$,
and elevation $z$

\[
\Phi(\phi,\lambda,z)=\intop_{0}^{z}g(\phi,\lambda,z')dz'
\]
where $g$ is the gravity acceleration. The \textit{geopotential height}
is the geopotential normalized to the standard gravity at mean sea
level $g_{0}=9.80665\:ms^{-2}$ 

\[
Z(\phi,\lambda,z)=\frac{\Phi(\phi,\lambda,z)}{g_{0}}
\]
In atmospheric physics it is natural to define the vertical coordinate
in terms of pressure $p$ rather than in terms of geometric height
$z$, assuming hydrostatic approximation (note that numerical climate
models always assume hydrostatic balance in their set of equations).
The geopotential height expressed in pressure coordinates behaves
as a streamfunction for the geostrophic wind vector, that is the first
order approximation of the wind vector for synoptic scale motions
at the midlatitudes (see \cite{hoskins_james_book_2014} for more
details). 

Contour plots of the anomalies of geopotential height in the mid-troposphere
at 500 hPa (about 5500 m) can be used to visualize the state of the
atmospheric circulation. Negative anomalies indicate low pressure
systems (\textit{cyclonic} anomalies, characterized by anticlockwise
circulation in the Northern hemisphere and typically bad weather),
while positive anomalies indicate high pressure systems \textit{(anticyclonic}
anomalies, characterized by clockwise circulation in the Northern
hemisphere and typically fair weather). As discussed in the main text,
heat waves are associated with persistent anticyclonic conditions,
as shown in figure 5(a).

We call $\mathbf{u}_{500}$ the velocity field vector at 500 hPa height.
The zonal (West-East) component is $u_{500}$ and the meridional (South-North)
component is $v_{500}$. Let us define the contribution to the specific
kinetic energy (the kinetic energy per unit mass) $KE_{500}=(u_{500}^{2}+v_{500}^{2})/2$.
For example in figure 1b) in the text we show for the Northern hemisphere
the long time average $\mathbb{E}\left[KE_{500}\right]$ from the
control run. Large values of $\mathbb{E}\left[KE_{500}\right]$ correspond
to areas of strong average zonal circulation in the mid-troposphere,
hence to the average position of the storm track.

\end{document}